\documentclass[3p]{elsarticle}
\usepackage{psfrag,amssymb}

\begin{document}

\def\bm#1{\mbox{\boldmath $#1$}} 
\newcommand{\comments}[1]{}
\newcommand{\ba}{\begin{eqnarray}}
\newcommand{\ea}{\end{eqnarray}}
\newcommand{\be}{\begin{equation}}
\newcommand{\ee}{\end{equation}}
\newcommand{\lan}{\langle}
\newcommand{\ran}{\rangle}
\newcommand{\grad}{\nabla}

\begin{frontmatter}

\title{A Fast Semi-implicit Method for Anisotropic Diffusion}

\author[ps]{Prateek~Sharma\corref{cor1}\fnref{fn1}}
\ead{psharma@astro.berkeley.edu}

\author[lbl]{Gregory W. Hammett}

\cortext[cor1]{Corresponding author}
\fntext[fn1]{Chandra Fellow}

\address[ps]{Theoretical Astrophysics Center and Astronomy Department, University of California, Berkeley, CA 94720}
\address[lbl]{Princeton Plasma Physics Laboratory, Princeton, NJ 08543}

\begin{abstract}

Simple finite differencing of the anisotropic diffusion equation, where diffusion is only along a given direction, does not ensure that the numerically
calculated heat fluxes are in the correct direction. This can lead to negative temperatures for the anisotropic thermal diffusion equation.
In a previous paper we proposed a monotonicity-preserving explicit method which uses limiters (analogous to those used in the solution of hyperbolic 
equations) to interpolate the temperature gradients at cell faces. However, being explicit, this method was limited by a restrictive 
Courant-Friedrichs-Lewy (CFL) stability timestep. Here we propose a fast, conservative, directionally-split, semi-implicit method which is second order accurate in space, is stable for large timesteps, and is easy to implement in parallel.
Although not strictly monotonicity-preserving, our method gives only small amplitude temperature oscillations
at large temperature gradients, and the  
oscillations are damped in time. With numerical experiments we show that our semi-implicit method
can achieve large speed-ups compared to the explicit method, without seriously violating the monotonicity constraint. This method can
also be applied to isotropic diffusion, both on regular and distorted meshes.
\end{abstract}

\begin{keyword}
implicit methods, finite differencing, monotonicity, anisotropic diffusion
\end{keyword}
\end{frontmatter}

\section{Introduction}

Anisotropic diffusion equation arises frequently in diverse applications: microscopic transport in magnetized plasmas \cite{bra65}; 
image processing \cite{per90}; diffusion-tensor magnetic resonance
imaging \cite{bas02}; thermal properties of crystals \cite{dia91};
transport in geological formations \cite{ber02}, etc. 
In \cite{sha07} we showed that simple finite-differencing of the anisotropic diffusion equation resulted in unphysical numerical heat fluxes, which lead to negative temperatures at large 
temperature gradients. Negative temperatures, in  addition to being unphysical, result in an imaginary sound speed and associated numerical instabilities.

Anisotropic diffusion equation satisfies important mathematical properties such as monotonicity preservation along the direction of diffusion, so that no new extrema are created in that direction and any existing extrema are not accentuated, i.e., maxima must drop or be unchanged, and minima must increase or be unchanged (e.g. see  \cite{ker81,nor07} and references therein).  For simplicity we will refer to this as an extrema reducing property.   In two or three dimensions the anisotropic diffusion equation of the form we are considering here can always be transformed to the form $\partial T(\alpha, \beta, t) / \partial t = \partial / \partial \alpha [ D(\alpha, \beta) \partial T / \partial \alpha]$, where $\alpha$ is a coordinate that varies along field lines and $\beta$ are field-line label coordinates that are constant along a magnetic field line, so the monotonicity properties of 1-D diffusion should be preserved.

In \cite{sha07} we proved that the use of slope limiters (e.g., see \cite{lev02}) to interpolate temperature gradients at cell faces guarantees that temperature 
extrema are not accentuated,
as required physically. 
The extrema-reducing property ensures that the temperature is positive for a CFL stable timestep.
Since the temperature is positive, numerical instabilities that plague simple finite differencing of anisotropic diffusion 
(because of an imaginary sound speed!), do not arise with the use of limiters. Because of this desirable property our method has been used in
astrophysical magnetohydrodynamic (MHD) simulations where thermal conduction is anisotropic, and large temperature gradients can arise \cite{sha10,par09,rus10}. In addition 
to thermal conduction, limiters have proved useful for anisotropic
viscosity with large gradients \cite{don09}, and may help in various
problems with anisotropic transport; e.g., cosmic ray streaming
\cite{sha09}.  There is another explicit method that has been applied
to astrophysical problems and avoids the problem of negative
temperature in presence of large temperature gradients, though it
is non-conservative and somewhat 
more complex \cite{ras08}. We also discuss some other recent work below.

A key limitation  of all explicit methods is that the timestep is
limited by the usual CFL condition, $\Delta t < \Delta x^2/2\chi_\parallel$, 
where $\Delta x$ is the grid spacing and $\chi_\parallel$ is the anisotropic diffusion coefficient. For some applications, this timestep constraint is rather severe and the conduction
timestep can be much smaller than the MHD CFL time limit. In such cases an implicit method, where there is no stability limit on the diffusion 
timestep, is desirable. Although it is straightforward to difference a
linear anisotropic diffusion equation implicitly, 
the resulting scheme is still not monotonicity-preserving. 
E.g., see Table 1 in \cite{bal08}, which shows that temperature oscillations 
remain till late times even with an implicit method, 
just as with the explicit schemes without limiters. One can try to solve the anisotropic diffusion equation fully implicitly, using limiters to prevent temperature
oscillations, but the nonlinearities in the limiters will require a careful iterative
treatment (some studies have found that these kinds of nonlinear
limiters make iterative solvers more difficult).
 
We have experimented with a Jacobian-free 
nonlinear iterative implicit method (a two-stage Richardson iteration
extension of the LGMRES(1,1) version of Loose GMRES \cite{Baker05},
a variant of the Generalized Minimal RESidual method\cite{saa86} with
restarting) but found that it requires a fairly large number of
iterations per time step, because the Jacobian matrix is not
strongly diagonally dominant for large timesteps and has a large
condition number.\footnote{Even for a linear 2-D Poisson problem on an $N \times N$ grid,
Loose GMRES or conjugate-gradient methods require ${\cal O}(N)$
iterations by themselves, or ${\cal O}(N^{1/2})$ iterations if combined
with a sufficiently good preconditioner like Modified
ILU\cite{Gustafsson78}.  On the other hand, the 
splitting method employed in this paper uses tri-dagonal implicit
solvers that are equivalent to only ${\cal O}(1)$ iterations and so are
quite fast by comparison.}
The fast method
proposed here might be able to serve as an effective preconditioner to
further accelerate an unsplit iterative method.

We have also experimented with an explicit method which is stable for timesteps longer than the CFL limit, 
and where the internal iteration time-steps are chosen based on the
properties of Chebyshev polynomials \cite{ale96}. We were not able to obtain a speed-up of more 
than $\sim$10 compared to the CFL-limited scheme, irrespective of resolution, for any of the parameters that we varied. Moreover, the parameters
for which maximum speed-up is obtained, without becoming numerically unstable, are difficult to choose 
(this is true even for isotropic diffusion!). 

Here we present a conservative, directionally-split, semi-implicit method which is numerically stable for any choice of timestep, and is easy to implement. The method is based on directional splitting where the heat fluxes in each direction are updated sequentially. 
The heat flux in each direction, e.g., $q_x=-\chi_\parallel b_x ({\vec b} \cdot \vec{\nabla})T$ (see Eq. \ref{eq:heat_flux}), consists of two terms: $-\chi_\parallel b_x^2 \partial T/\partial x $, the `normal' term where temperature gradient need not be interpolated, and $-\chi_\parallel b_x b_y \partial T/\partial y$, the `transverse'  term which involves temperature interpolation with limiters. In our directionally-split method the `normal' terms are treated implicitly, and the transverse terms are treated explicitly.
The directional splitting of `normal' implicit terms results in a tridiagonal matrix which can be solved very quickly. The explicit treatment of `transverse' terms with limiters ensures that extrema are not accentuated. The resulting scheme, while not strictly monotonicity-preserving for large timesteps, results
in only small amplitude temperature oscillations which are damped in time. Speed-up of order 100-1000, compared to the explicit scheme, is easily achieved for our test problems.

There has been some interesting recent work on another approach to the
problem of preserving positivity in presence of anisotropic
diffusion tensors (or diffusion on distorted meshes), based on
expressing the flux at cell faces in terms of the advected quantity at
the cell centers on either side of the face (a ``two-point flux
expression''), but where the coefficients of this flux depend on the
transverse gradients and so is nonlinear (for example see
\cite{Lipnikov10, Sheng09}).  
Future work could compare the 
nonlinear limiters and implicit solvers that are used in our algorithm with these
other algorithms on the types of test problems considered here and in
\cite{sha07}.

The paper is organized as follows: Section 2 presents the method in detail and shows that the scheme is linearly stable for large timesteps. Section 3 presents results from three
test problems which show the practical utility of our method. We conclude and discuss applications of our method in Section 4. 

\section{The Method}
The anisotropic diffusion equation in its simplest form is given by
\ba 
\label{eq:aniso_cond}
\frac{\partial T}{\partial t} &=& - \vec{\nabla} \cdot \vec{q}, \\
\label{eq:heat_flux}
\vec{q} &=& - \chi_\parallel \vec{b} (\vec{b}\cdot \vec{\nabla}) T = - \chi_\parallel \vec{b}
\nabla_\parallel T, 
\ea 
where $T$ is the temperature, $\vec{q}$ is the heat flux along magnetic field lines,
$\chi_\parallel$ is the thermal diffusion coefficient~(with dimensions $L^2T^{-1}$), 
and $\vec{b}$ is the magnetic field unit vector. We will assume $\vec{b}$ as a given function
of space and time, and $\chi_\parallel$ as being a constant for simplicity (in \cite{sha07} we showed 
that a harmonic average should be used for interpolation of the diffusivity for numerical stability). We assume two-dimensions 
with a uniform Cartesian grid and a constant diffusion coefficient. The generalization to 
a nonlinear diffusivity, general coordinate system, and three dimensions is straightforward.

The semi-implicit method is obtained by directional splitting of the heat flux updates in each direction, given by
\ba
\nonumber
\frac{T_{i,j}^{\star}-T_{i,j}^n}{\chi_\parallel \Delta t} &=& b_{x,i+1/2,j}^2 \frac{T_{i+1,j}^{\star}-T_{i,j}^{\star}}{\Delta x^2} 
- b_{x,i-1/2,j}^2 \frac{T_{i,j}^{\star}-T_{i-1,j}^{\star}}{\Delta x^2} \\
\label{eq:xup}
&+& \frac{b_{x,i+1/2,j} b_{y,i+1/2,j}}{\Delta x \Delta y} \overline{\Delta T}_{i+1/2,j}^n- \frac{b_{x,i-1/2,j} b_{y,i-1/2,j}}{\Delta x\Delta y} \overline{\Delta T}_{i-1/2,j}^n, \\
\nonumber
\frac{T_{i,j}^{n+1}-T_{i,j}^\star}{\chi_\parallel \Delta t} &=& b_{y,i,j+1/2}^2 \frac{T_{i,j+1}^{n+1}-T_{i,j}^{n+1}}{\Delta y^2} 
- b_{y,i,j-1/2}^2 \frac{T_{i,j}^{n+1}-T_{i,j-1}^{n+1}}{\Delta y^2} \\
\label{eq:yup}
&+& \frac{b_{y,i,j+1/2} b_{x,i,j+1/2}}{\Delta x \Delta y} \overline{\Delta T}_{i,j+1/2}^\star- \frac{b_{y,i,j-1/2} b_{x,i,j-1/2}}{\Delta x \Delta y} \overline{\Delta T}_{i,j-1/2}^\star,
\ea
where magnetic field unit vectors are interpolated at the appropriate cell faces (simple averaging is fine for magnetic field unit vectors), and  
\ba
\overline {\Delta T}_{i+1/2,j} &=& L\left ( T_{i+1,j+1} - T_{i+1,j}, T_{i+1,j}-T_{i+1,j-1}, T_{i,j+1}-T_{i,j}, T_{i,j}-T_{i,j-1} \right ), \\
 \overline {\Delta T}_{i,j+1/2} &=& L \left ( T_{i+1,j+1} - T_{i,j+1}, T_{i,j+1}-T_{i-1,j+1}, T_{i+1,j}-T_{i,j}, T_{i,j}-T_{i-1,j} \right ),
\ea
are the temperature differences centered at appropriate faces, and $L$ stands for a limiter. See \cite{sha07} for a discussion of limiters in this context; 
here we will use the slope limiter of van Leer \cite{van77,lev02}; $L(a,b,c,d)=L(L(a,b),L(c,d))$ is symmetric in its arguments, where 
\ba
\nonumber
L(a,b) &=& \frac{2ab}{a+b}, \hspace{1in} {\rm if}~ab>0, \\ 
\label{eq:vanLeer}
&=& 0, \hspace{1.25in} {\rm otherwise}.
\ea
We have experimented with other slope limiters (e.g., minmod and monotonized central [MC] limiters). We observed that monotonicity properties are much better with diffusive limiters (such as minmod and van-Leer) as compared to sharper limiters such as the MC limiter (see \cite{lev02} for properties of different limiters) for our semi-implicit scheme with large timesteps. However, more diffusive limiters result in larger perpendicular diffusion.

Defining the components of the diffusion operator on the right hand side
of Eq. (\ref{eq:aniso_cond}) as 
\be
\label{eq:operator}
{\cal D}_{ij} = - \frac{\partial}{ \partial x_i} \left ( \chi_\parallel b_i b_j \frac{\partial}{\partial x_j} \right )
\ee
(there is no implied summation
on the right hand side of this definition), then the method in
Eqs. (\ref{eq:xup}) \& (\ref{eq:yup}) can be expressed as
\be
\label{eq:op_2step}
T^{n+1} = 
(1+\Delta t {\cal D}_{yy})^{-1} 
(1-\Delta t {\cal D}_{yx}) 
(1+\Delta t {\cal D}_{xx})^{-1}
(1 - \Delta t {\cal D}_{xy}) T^{n}.
\ee

Our formulation treats the `normal' temperature derivative terms, which are guaranteed to result in a heat flux in the correct direction, 
implicitly. The `transverse' terms are treated explicitly, and employ limiters that ensure that the temperature extrema are not accentuated.
Directional splitting results in a quickly solvable tridiagonal matrix for each directional update. Instead of updating $q_x$ followed by $q_y$ we can also
update the heat fluxes in the reverse order. Results do not depend
substantially on the order of updates. Since our split operators are
individually only first order accurate in 
time, Strang-splitting (e.g., see \cite{lev02}) will not improve the
accuracy of our method (and high accuracy is not a priority for 
components of the solution that are strongly damped anyway).
Moreover, Strang-splitting applied to Eqs. (\ref{eq:xup}) and
(\ref{eq:yup}) is numerically unstable for large timesteps.

\subsection{Linear Stability Analysis}
\label{sec:LSA}
\begin{figure}
\centering
\includegraphics[width=4in,height=3.5in]{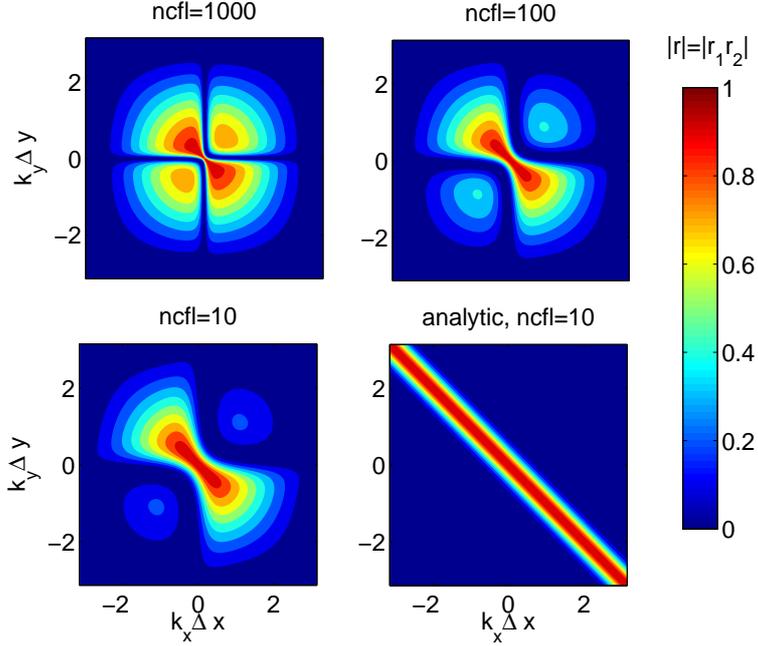}
\caption{Contour plots of the amplification factor $|r|=|r_1r_2|$ (see Eqs. \ref{eq:amp1}, \ref{eq:amp2}) for $b_x=b_y=1/\sqrt{2}$ and different Courant factors (ncfl, see Eq. \ref{eq:ncfl}). Also shown is the analytic amplification factor ($e^{-k_\parallel^2 \Delta x^2 {\rm ncfl}/4}$, assuming $\Delta x=\Delta y$, where $k_\parallel = k_x b_x+k_y b_y$) for ncfl=10; there is no damping along $k_x=-k_y$ because temperature gradient (which is 
along $\vec{k}$) is perpendicular to $\vec{b}$. 
\label{fig:amp}}
\end{figure}

One can perform the von Neumann linear stability analysis on Eqs. (\ref{eq:xup}) and (\ref{eq:yup}). Let us assume a single temperature mode 
$T(x,y,t)=T_0r(t)e^{-i (k_x x + k_y y)}$, where $r(t)$ is the amplification factor in time, and $k_x$, $k_y$ are the wavenumbers in the 
$x-$ and $y-$ directions. The amplification factor can be written as $r=r_1r_2$, where $r_1$ and $r_2$ are amplification factors for the substages in Eqs.
(\ref{eq:xup}) and (\ref{eq:yup}), respectively. On substituting the discretized temperature eigenmode in Eq. (\ref{eq:xup}), and using trigonometric identities, one obtains
\be
\label{eq:amp1}
r_1 = \frac{1- \frac{\chi_\parallel \Delta t}{\Delta x \Delta y} b_x b_y \sin(k_x \Delta x) \sin(k_y \Delta y) }{1 + 4 \frac{\chi_\parallel \Delta t}{\Delta x^2} b_x^2 \sin^2(k_x\Delta x/2)},
\ee
where, for simplicity, we have assumed that $b_x$, $b_y$ are constant in
space. We use arithmetic averaging instead of the nonlinear limited
averaging for the transverse temperature gradient. 
Similarly, for update in the $y-$ direction,
\be
\label{eq:amp2}
r_2 = \frac{1- \frac{\chi_\parallel \Delta t}{\Delta x \Delta y} b_x b_y \sin(k_x \Delta x) \sin(k_y \Delta y) }{1 + 4 \frac{\chi_\parallel \Delta t}{\Delta y^2} b_y^2 \sin^2(k_y\Delta y/2)}.
\ee
Now the amplification factor after a full timestep $r=r_1r_2$ is
\be
\label{eq:amp_sym}
r = \frac{\left [1-A_x A_y \cos(k_x \Delta x/2) \cos(k_y \Delta y/2) \right ]^2}{(1+A_x^2)(1+A_y^2)},
\ee
where $A_x=2 \sqrt{\chi_\parallel \Delta t} \frac{b_x}{\Delta x} \sin(k_x\Delta x/2)$ and $A_y=2 \sqrt{\chi_\parallel \Delta t} \frac{b_y}{\Delta y} \sin(k_y\Delta y/2)$. From Eq. (\ref{eq:amp_sym}) we get
$$
r \leq \frac{(1 + |A_x A_y|)^2}{1+A_x^2A_y^2+A_x^2+A_y^2} = \frac{1 + A_x^2 A_y^2 + 2 |A_xA_y| }{1+A_x^2A_y^2+A_x^2+A_y^2},
$$
which is guaranteed to be $\leq 1$ since $2 |A_x A_y| \leq A_x^2 + A_y^2$ for all real $A_x$, $A_y$. Thus our scheme (Eqs. \ref{eq:xup} \& \ref{eq:yup}) is unconditionally stable like the usual implicit methods (see Fig. \ref{fig:amp}). The unconditional stability of our scheme also holds for the case of a general symmetric, positive diffusion tensor
$$
D = \left ( \begin{array}{cc}
d_{xx} & d_{xy} \\
d_{xy} & d_{yy} \end{array} \right ),
$$
with non-negative eigenvalues (i.e., $d_{xx}d_{yy} \ge d_{xy}^2$). Although large timesteps are stable, we cannot use very large timesteps because of loss of accuracy.

 Fig. \ref{fig:amp} shows the amplification factor ($|r|=|r_1r_2|$) for a timestep longer than the Courant time step (characterized by ncfl; see Eq. \ref{eq:ncfl})
 for a fixed magnetic field unit vector $b_x=b_y=1/\sqrt{2}$. As expected, the amplification factor is close to the analytic expectation (also shown in the same figure) for smaller ncfl. The damping rate is not sufficiently large (compared to the analytic solution) for unit vectors corresponding to the first and the third quadrant in $k$-space. The maximum growth rate for the points in the first and third quadrants occurs at small scales (compared to the box size), but this 
 scale and the amplification factor becomes larger for a larger ncfl. Similar low damping rate arises for the Crank-Nicolson method for isotropic diffusion for large time steps. The slow damping of modes parallel to $\vec{b}$ at small scales
 can be avoided if we modify our scheme to a 4-step scheme, with  the implicit terms in Eqs. (\ref{eq:xup}) \& (\ref{eq:yup}) only applied for $\Delta t/2$ and two extra fully-implicit steps applied for $\Delta t/2$. Schematically this 4-step scheme is given by (see Eq. \ref{eq:operator} for notation)
 \be
 \label{eq:4step}
 T^{n+1} = 
(1+\Delta t {\cal D}_{yy}/2)^{-2} 
(1-\Delta t {\cal D}_{yx}) 
(1+\Delta t {\cal D}_{xx}/2)^{-2}
(1 - \Delta t {\cal D}_{xy}) T^{n}.
 \ee
However, for the test problems discussed in \S \ref{sec:test} our 2-step method behaves satisfactorily even for ncfl as large as 1000. And moreover, perpendicular diffusion, non-monotonicity, and computational cost are slightly worse for the 4-step method (one can try different orderings of the operators 
in Eq. \ref{eq:4step} but this, and its analog where $x$ and $y$ are interchanged, worked best for our test problems). Thus we do not discuss the 4-step method in detail.
 Another feature in Fig. \ref{fig:amp}, which is seen for all ncfl, is that the smallest scale modes in the direction perpendicular to field lines are damped; this corresponds to cross-field diffusion at small scales.
 
 Expanding the amplification factor ($r$; Eqs. \ref{eq:amp1} \& \ref{eq:amp2}) in the limit $k_x\Delta x$, $k_y\Delta y \ll 1$ (i.e., large length scales), 
 one gets
\be
\label{eq:amp}
 r = \frac{(1- \chi_\parallel k_x k_y b_x b_y \Delta t)^2 }{(1 + \chi_\parallel k_x^2 b_x^2 \Delta t)(1 + \chi_\parallel k_y^2 b_y^2 \Delta t)},
\ee
which should be compared to the analytic amplification factor $r_{\rm a} = e^{-k_\parallel^2 \chi_\parallel \Delta t}$, where $k_\parallel = k_x b_x + k_y b_y$.
In the limit $\chi_\parallel k_\parallel^2 \Delta t \ll 1$, the difference between the numerical and analytical amplification factors is ${\cal O}( [\chi_\parallel k_\parallel^2 \Delta t]^2)$; i.e., this method is first order in time.

It is instructive to do a similar stability analysis in three dimensions, with $r_1$, $r_2$, $r_3$ as the amplification factor in each directional update. In this
case, 
$$
r_1 = \frac{1- \frac{\chi_\parallel \Delta t}{\Delta x \Delta y} b_x b_y \sin(k_x \Delta x) \sin(k_y \Delta y) - \frac{\chi_\parallel \Delta t}{\Delta x \Delta z} b_x b_z \sin(k_x \Delta x) \sin(k_z \Delta z) }{1 + 4 \frac{\chi_\parallel \Delta t}{\Delta x^2} b_x^2 \sin^2(k_x\Delta x/2)},
$$
and analogous expressions are obtained for other directions. It is easy to show that the absolute value of the amplification factor $|r| = |r_1r_2r_3|$ is not
guaranteed to be $\leq 1$ for a large $\Delta t$. Thus, our scheme is not unconditionally stable in three dimensions. By numerically evaluating $|r|$ for different parameters\footnote{We calculate the maximum of $|r|=|r_1r_2r_3|$ on a grid with resolution upto $320^3$ in $(k_x\Delta x, k_y\Delta y, k_z\Delta z)$, for a given $(\chi_\parallel \Delta t b_x^2/\Delta x^2, \chi_\parallel \Delta t b_y^2/\Delta y^2, \chi_\parallel \Delta t b_z^2/\Delta z^2)$. Then we try to find the maximum of $\chi_\parallel \Delta t /\Delta x^2$ (assuming $\Delta x = \Delta y = \Delta z$) for which $|r| \leq 1$.} we have verified that the stability condition for our scheme in three dimensions is (assuming $\Delta x=\Delta y=\Delta z$)
$$
\frac{\chi_\parallel \Delta t}{\Delta x^2} \leq 8.25.
$$
The corresponding stability limit for the explicit scheme in three dimensions is $\chi_\parallel \Delta t/\Delta x^2 \leq 0.444$. Thus, our scheme can attain a 
speed-up of $\approx 18$ relative to the explicit method. Although our 2-step scheme (Eq. \ref{eq:op_2step}) is not unconditionally stable in three dimensions, we have numerically verified that the 4-step method (Eq. \ref{eq:4step}) is unconditionally stable in three dimensions.

We have also experimented with a variant of the alternate direction
implicit (ADI) schemes for Eq. (\ref{eq:aniso_cond}), inspired by their
application to isotropic diffusion \cite{pre92}.  Specifically, we tried (see Eq. \ref{eq:operator} for notation)
$$T^{n+1} = 
(1+\Delta t {\cal D}_{yy}/2)^{-1} 
(1 - \Delta t [ {\cal D}_{yx} +{\cal D}_{xy} + {\cal D}_{xx}]/2 )
(1+\Delta t {\cal D}_{xx}/2)^{-1}
(1 - \Delta t [{\cal D}_{xy} +{\cal D}_{yx} + {\cal D}_{yy}]/2 ) T^{n}.$$
However,  
presence of the transverse terms ($\partial^2T/\partial x\partial y$) makes the scheme unstable for timesteps larger than a few times the CFL timestep, unlike in the case of isotropic diffusion where it is unconditionally stable. Even for isotropic diffusion, ADI does not give strong damping in the
large time step limit, i.e., it is A-stable but not L-stable. Therefore, we do not consider ADI further.

Notice that our fully implicit scheme is only first order accurate in
time, but for dissipative processes this is often adequate, since one is
most interested in well-resolved components of the solution which are
only weakly damped in a single time step.
Another variant of Eqs. (\ref{eq:xup}) and (\ref{eq:yup}), where the explicit `transverse'  ($\partial^2 T/\partial x \partial y$) terms are symmetrized with respect to the $x-$ and $y-$ updates (we were trying this to see if this scheme has better monotonicity properties as compared to our method), is numerically unstable for large $\Delta t$s 
even though  the linear stability analysis predicts an unconditional stability. Thus, linear stability is only a necessary (and not sufficient) condition for numerical stability, especially since the limiters are nonlinear.

\section{Numerical Tests}
\label{sec:test}

In this section we describe various tests for our semi-implicit scheme (Eqs. \ref{eq:xup} and \ref{eq:yup}).

\subsection{Diffusion in a ring}
\label{sec:ring}
The ring diffusion test (see \cite{par05,sha07}) involves the diffusion of a hot patch in fixed circular magnetic field lines. This is a crucial test to check
monotonicity properties of the anisotropic diffusion scheme because field lines make all possible angles with respect to the Cartesian grid. At late times
 (a few diffusion times across the ring), the temperature is expected to be uniform along each magnetic field line in the ring. The computational domain is a $[-1,1]\times[-1,1]$
 Cartesian box. The initial temperature distribution is
 \ba
 \nonumber
 T &=& 10 \hspace{1in} {\rm if}~.5<r<0.7~{\rm and}~\frac{11}{12}\pi<\theta<\frac{13}{12}\pi, \\
    &=& 0.1\hspace{1in} {\rm otherwise},
 \ea
where $r=\sqrt{x^2+y^2}$ and $\tan\theta=y/x$, and $b_x=-y/\sqrt{x^2+y^2}$, $b_y=x/\sqrt{x^2+y^2}$. Reflective boundary condition ($\partial T/\partial x=0$ at boundaries in the $x-$ direction, $\partial T/\partial y = 0$ at $y-$ boundaries) is 
used for temperature; magnetic field and conduction vanishes outside $r = 1$. The parallel conduction coefficient $\chi_\parallel = 0.01$; there is no explicit 
perpendicular diffusion. 

\begin{figure}
\centering
\includegraphics[width=6in,height=2.5in]{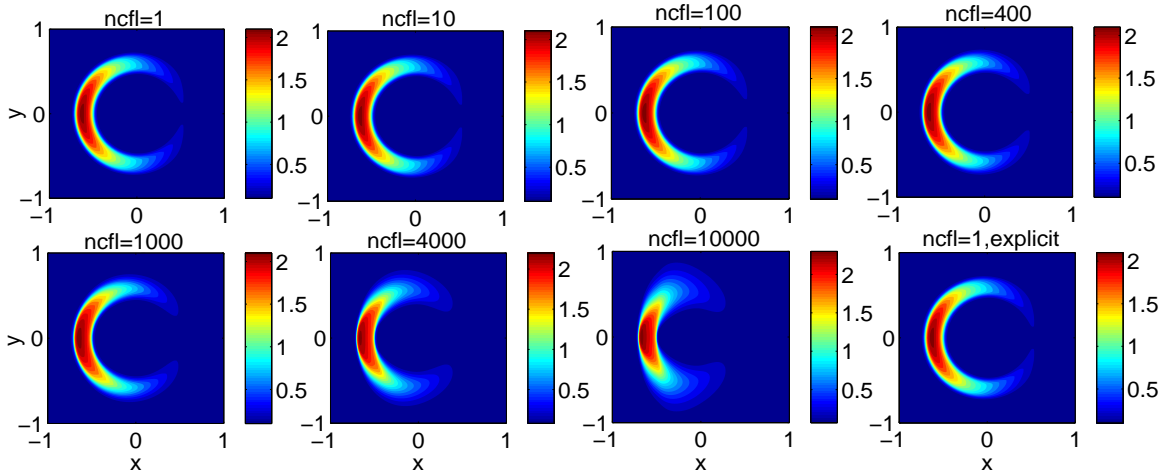}
\caption{Temperature contour plots at $t=20$ for the ring diffusion test problem using a $512\times 512$ grid. Our semi-implicit method (Eqs. \ref{eq:xup} and \ref{eq:yup}) is used with different CFL numbers (ncfl). Also shown is the temperature plot with the fully explicit method (using the van Leer limiter) 
for comparison.
\label{fig:diff_cfl_512}}
\end{figure}

We quantify the timestep by ncfl, where 
\be
\label{eq:ncfl}
\Delta t={\rm ncfl} \Delta x^2/4\chi_\parallel. 
\ee
Since our scheme is unconditionally stable in two dimensions, we experiment with the CFL number (ncfl) to quantify the speed-up relative to the explicit method that we can obtain, without degrading the solution. Fig. \ref{fig:diff_cfl_512} shows the  temperature contour plots at $t=20$ using different ncfl for a $512\times512$ box. As expected from section \ref{sec:LSA}, our scheme is numerically stable 
even for ncfl$\gg$ 1. However, the solution deteriorates for an extremely large ncfl; e.g., the temperature profile for ncfl=10000 looks quite different from
rest of the others as there is considerable numerical diffusion out of the circular ring. This figure shows that large speed-ups ($\sim$ 1000 in the case of Fig. \ref{fig:diff_cfl_512}) are possible as compared to the explicit method. Moreover, temperature oscillations at extrema are not as severe as schemes without
limiters.

Extrema are not accentuated for our semi-implicit method because the transverse term, responsible for non-monotonicity, vanishes at temperature extrema
because of limited averaging.
However, it is not guaranteed that the temperature will be bound by the initial temperature extrema; this is because temperature oscillations may arise at
non-extremal locations for a large CFL factor (ncfl). These newly created extrema will not be accentuated though. We never encountered temperature oscillations for the fully-explicit method with limiters which uses a CFL-stable timestep.

\begin{figure}
\centering
\includegraphics[width=5in,height=2.5in]{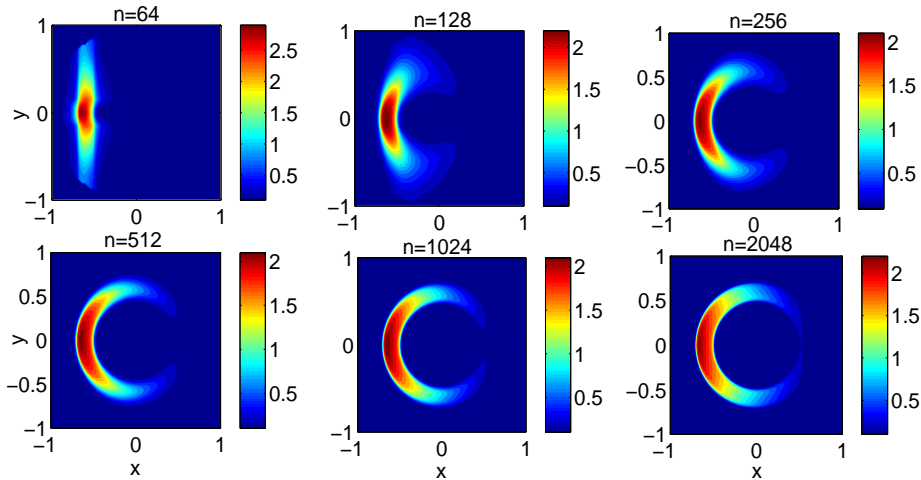}
\caption{Temperature contour plots at $t=20$ for different grid resolutions but a fixed CFL number (ncfl=1000), using our 
semi-implicit method.
\label{fig:cfl_100}}
\end{figure}

Fig. \ref{fig:cfl_100} shows the temperature profiles at $t=20$ for different grid resolutions but with a fixed ncfl=1000. The figure shows that the temperature
profiles are very similar for $n\geq 512$. 
The maximum speedup relative to the explicit method (i.e., maximum
value of ncfl), without seriously affecting the solution, is achieved for
the highest resolution simulations (where significant speed-up is
in-fact desired), as seen from Figs. \ref{fig:diff_cfl_512} \&  
\ref{fig:cfl_100}. 
While ncfl=1000 can be used for $n=512$, the solution for 
a lower grid resolution with ncfl=1000 is quite diffusive in the perpendicular direction;
also parallel diffusion seems to be suppressed (as is the case for ncfl=10000 and $n=512$ in Fig. \ref{fig:diff_cfl_512}).

\begin{figure}
\centering
\includegraphics[width=3in,height=3in]{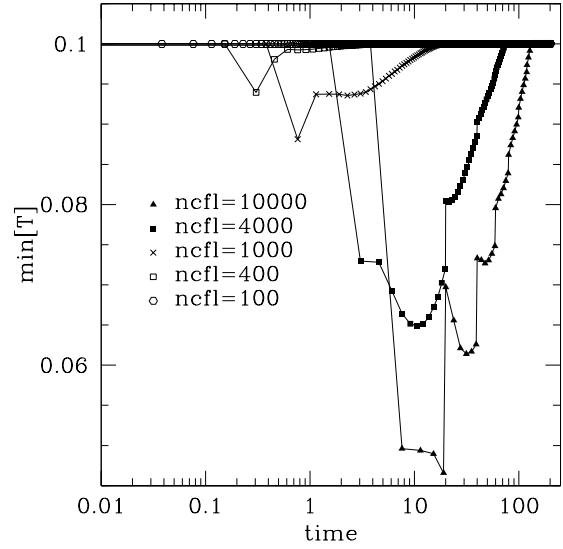}
\caption{Minimum temperature over the computational domain as a function of time (all timesteps are plotted) for our semi-implicit method using 
different Courant factors (ncfl). As in Fig. \ref{fig:diff_cfl_512}, grid resolution is $512\times512$. For comparison, the minimum temperature for the 
explicit method without limiters and with ncfl=1 is -0.41.
\label{fig:tmin}}
\end{figure}

As mentioned earlier, our semi-implicit scheme is not guaranteed to be monotonicity-preserving. However, the oscillations originating at large temperature
gradients are of small amplitude and are damped away quickly in time (see Fig. \ref{fig:tmin}). In contrast, temperature oscillations for the explicit methods without limiters are large and persist till late times (see Fig. 7 in \cite{sha07}). Of course, the amplitude of temperature oscillations is proportional to 
the ratio of maximum to minimum temperature at the discontinuity, but Fig. \ref{fig:tmin} shows that the temperature is still maintained positive for ncfl as large as 10000! The temperature ratio of 100 (as in our test problem) is similar to the temperature range found in practical applications such as the transition of the solar chromosphere at $10^4$ K to the coronal temperature of $10^6$ K. Notice that the minimum temperature respects monotonicity constraint for the 
first timestep because all points in the initial condition are extrema and the transverse term vanishes. The temperature oscillations start from non-extremal points at later times.

\begin{figure}
\centering
\includegraphics[width=3in,height=3in]{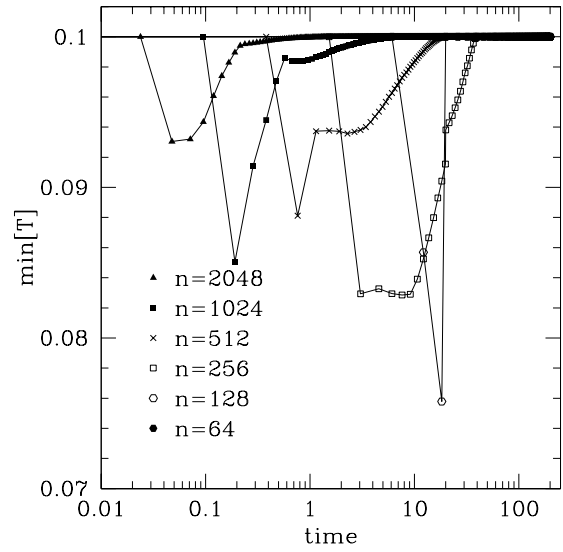}
\caption{Minimum temperature over the computational domain as a function of time (all timesteps are plotted) for different resolutions using our semi-implicit method. As in Fig. \ref{fig:cfl_100}, the Courant factor (ncfl) is fixed to be 1000.
\label{fig:tmin_nx}}
\end{figure}

Fig. \ref{fig:tmin_nx} shows the minimum temperature over the domain as a function of time for different grid resolutions and a fixed ncfl=1000. While Fig. \ref{fig:tmin} shows 
a clear trend of increased non-monotonicity  as ncfl is increased, there is no systematic variation in the magnitude of temperature oscillations with the 
grid resolution for a fixed ncfl. This is expected because the factor $ \chi_\parallel \Delta t/\Delta x^2$ is of the same order ($\sim$ncfl) for all resolutions.
However, Fig. \ref{fig:cfl_100} clearly shows that  for a fixed ncfl a more accurate temperature profile is obtained for a higher grid resolution.

For a realistic problem ncfl should be $\lesssim (l/\Delta x)^2$, where $l$ is the scale on which we want temperature to be calculated accurately. Notice, that this factor increases with the grid resolution for a fixed $l$, so higher resolution runs will more accurate
for a fixed ncfl (see Fig. \ref{fig:cfl_100}). 
 Another constraint on ncfl comes from the positivity requirement. The magnitude of non-monotonicity is roughly independent of the grid resolution for a fixed ncfl (see Fig. \ref{fig:tmin_nx}), but depends on the ratio of maximum to minimum temperature at the discontinuity. E.g., for our test problem the initial temperature ratio is 100 and the maximum relative non-monotonicity (defined as 
$\{T_{{\rm min},0}-{\rm min}[T]\}/T_{{\rm min},0}$, where $T_{{\rm min},0}$ is the initial minimum temperature and ${\rm min}[T]$ is the minimum 
temperature of all times) for ncfl=1000 
(see Fig. \ref{fig:tmin_nx}) is $\approx (.1-.08)/.1=0.2$. We have numerically verified that the relative non-monotonicity 
scales with the maximum to minimum temperature ratio (and of course non-monotonicity is larger for a larger ncfl). As mentioned before, non-monotonicity 
is worse for steeper limiters such as the MC limiter, but is less severe for diffusive limiters such as minmod.

\subsection{Convergence \& measuring $\chi_{\perp, {\rm num}}$}
We perform a test problem with a smooth solution, described in \cite{sov04}, to measure perpendicular numerical diffusion as a function of grid resolution and the Courant factor (ncfl). A 
two-dimensional Cartesian box ([-0.5,0.5]$\times$[-0.5,0.5]) is initialized with a zero temperature. Temperature is fixed to be zero at the domain boundaries at all times. We solve the anisotropic diffusion equation (Eq. \ref{eq:aniso_cond}) with a source term
\be
\frac{\partial T}{\partial t} = - \vec{\nabla} \cdot \vec{q} + Q,
\ee
where $Q = 2\pi^2 \cos\pi x \cos \pi y $. The fixed magnetic field is generated by a flux function $\phi \propto \cos \pi x \cos \pi y$, so that the magnetic field unit vectors are along the contours of constant $Q$. And since temperature is driven by the source term, temperature is always constant along field lines. 
If there is
no diffusion across field lines, temperature should rise with time. However, because of finite numerical diffusion in the perpendicular direction, it reaches a
steady state. The steady state solution for the temperature, if we assume a finite perpendicular diffusivity $\chi_\perp$, is $T = \chi^{-1}_\perp  \cos \pi x \cos \pi y$, independent of $\chi_\parallel$. We use the asymptotic value (in time) 
of the maximum temperature to
calculate $\chi_{\perp, {\rm num}}=1/T(0,0)$. This test is slightly modified from \cite{sov04} in that we do not include an explicit perpendicular diffusivity; this is
because for the problems of our interest perpendicular conduction is negligible.

\begin{figure}
\centering
\includegraphics[width=3in,height=3in]{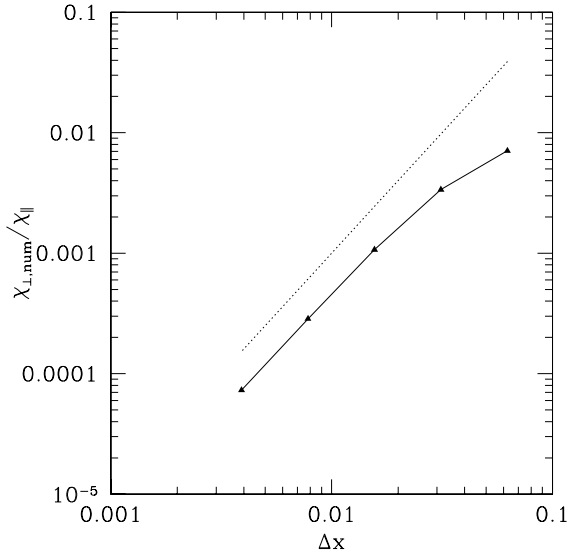}
\caption{The ratio of perpendicular (numerical) to parallel diffusivity as a function of grid size (triangles) for the smooth test problem in \cite{sov04}. The Courant factor (ncfl) is fixed to be 1000. Dotted line shows a second order convergence. \label{fig:conv}}
\end{figure}

Fig. \ref{fig:conv} shows the ratio of the perpendicular numerical diffusivity and the parallel diffusivity ($\chi_{\perp, {\rm num}}/\chi_\parallel$) as a function of grid resolution for
a fixed ncfl=1000. Perpendicular diffusion, which scales with $\chi_\parallel$, shows close to a second order convergence with the grid resolution, as expected.
Numerical diffusion is not sensitive to the Courant factor (ncfl); e.g., we verified that $\chi_{\perp,{\rm num}}/\chi_\parallel$  is roughly independent of ncfl up to ncfl=10000 for  $n=256$. Close to second order convergence of perpendicular numerical diffusion (and the independence of ncfl) was also seen for the 
ring diffusion test problem in section \ref{sec:ring}.

\subsection{Thermal Instability}
\begin{figure}
\centering
\psfrag{A}{ncfl=$\Delta t/\Delta t_{\rm exp}$}
\psfrag{B}{ncfl=1}
\psfrag{C}{$\log_{10}T$}
\includegraphics[width=5.8in,height=3in]{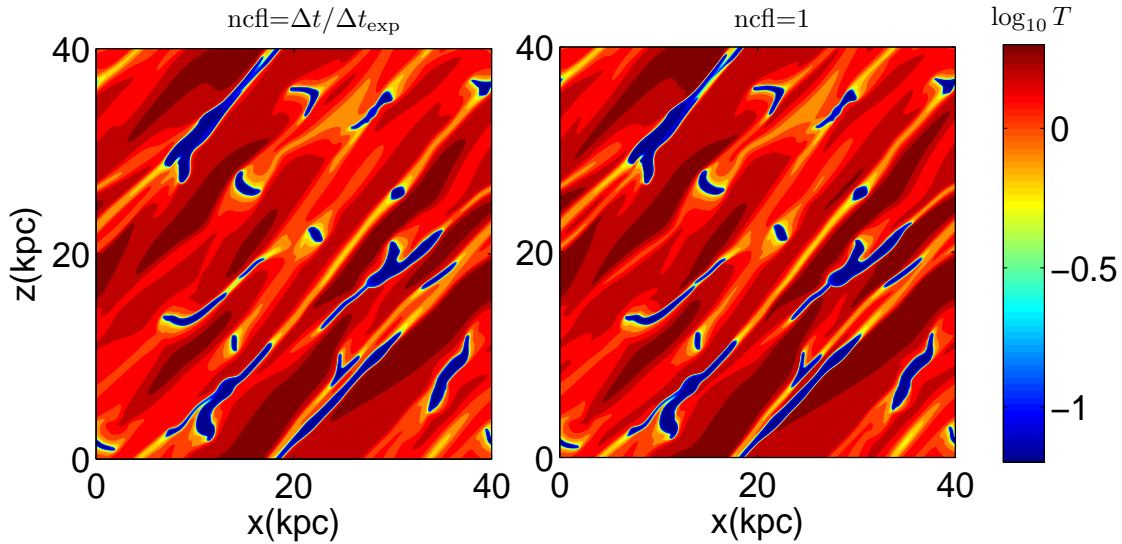}
\caption{$Log_{10}$ temperature (keV) at $0.95$ Gyr ($\approx$ 10 cooling times in the initial state) using our semi-implicit method (left) and the explicit method (right). The timestep for the conduction module in the semi-implicit scheme is chosen to be equal to the CFL timestep for the rest of the code ($\Delta t$) so that ncfl=$\Delta t/\Delta t_{\rm exp}$, where $\Delta t_{\rm exp} = \Delta x^2/4\chi_{\parallel,{\rm max}}$, and $\chi_{\parallel, {\rm max}}$ is the maximum thermal diffusivity over the whole box. The
conduction module is subcycled $\approx \Delta t/\Delta t_{\rm exp}$ times for the explicit method.
\label{fig:TI}}
\end{figure}

We also tested our method for a realistic astrophysical application, namely thermal instability in the intracluster medium, the X-ray emitting hot plasma pervading the massive clusters of galaxies. For astrophysical motivation and details about numerical set up see \cite{sha10}.  We perform
two-dimensional MHD simulations with anisotropic thermal conduction, using a 
periodic Cartesian box (40 kpc $\times$ 40 kpc) with 1024 grid points in each direction. The initial temperature
is 0.78 keV and initial electron number density is 0.1 cm$^{-3}$. Identical pattern of small amplitude density/temperature perturbations are initialized 
(such that the pressure is uniform) to seed the thermal instability. The functional form of heating and cooling is such that cooling increases faster 
than heating for the cooler plasma, and vice versa for the hotter plasma. Thus, heating/cooling runs away and the plasma segregates into a two-phase
medium. Net heating averaged over the whole box equals net cooling, so that the total thermal energy content of the box does not change with time. Thermal instability is
in the isobaric limit; i.e., cooling time $\gtrsim$ sound crossing time over all scales. Thermal conduction is primarily along field lines aligned initially at $45^0$ to the box; field lines roughly maintain their geometry even in the nonlinear stage. Small diffusion perpendicular to field lines is added for numerical convergence (see \cite{sha10} for details). Cold filaments aligned along the direction of the local magnetic field arise nonlinearly (see Fig. \ref{fig:TI}) 
because thermal conduction along field lines suppresses growth of small scale modes.

Fig. \ref{fig:TI} shows the temperature in the nonlinear state of thermal instability obtained by treating thermal conduction using our semi-implicit scheme 
(left) and using the explicit scheme (right) with the van Leer limiter. The temperature plots are almost identical, establishing the practical utility of our
method. The conduction timestep ($\Delta t_{\rm exp} = \Delta x^2/4\chi_{\parallel,{\rm max}}$, where $\chi_{\parallel, {\rm max}}$ is the maximum thermal diffusivity over the whole box and $\Delta x=\Delta y$ is the grid size) in the initial state is 3 times the CFL timestep limit of rest of the code ($\Delta t$). As thermal 
instability becomes nonlinear and the hottest plasma becomes hotter, $\Delta t_{\rm exp}$ decreases rapidly relative to $\Delta t$, because of the sensitive dependence 
of conductivity on temperature ($\chi_\parallel \propto T^{5/2}$; see \cite{bra65}). The temperature-dependent conductivity is interpolated at the faces 
using the current temperature (see \cite{sha07} for details about interpolation of conductivity). 
At 0.95 Gyr (the time corresponding to Fig. \ref{fig:TI}) $\Delta t/\Delta t_{\rm exp} \approx 100$; thus the explicit method is subcycling the conduction module for 100 times, whereas the conduction module is applied only once for 
our semi-implicit scheme. Situation become worse with time because the hottest plasma in the box becomes hotter in time! Thus we are able to run much faster with 
our semi-implicit scheme, without affecting the solution and without violating temperature positivity. This example demonstrates the practical utility of our method. Also, recall that the stability limit for explicit diffusion scales as $\Delta x^2$ compared to $\Delta x$ for the hyperbolic terms, so our scheme will be
even more useful at higher resolution and with mesh refinement.

\section{Conclusions}

We present a simple, directionally-split, conservative, semi-implicit method for anisotropic diffusion which is linearly stable for large timesteps.
Directional splitting results in a tridiagonal matrix equation for each direction, which can be solved exactly and efficiently. For problem on a $N \times 
N$ grid our scheme (Eqs. \ref{eq:xup} \& \ref{eq:yup}) requires two independent tridiagonal solves. In comparison, the fastest unsplit methods like LGMRES will require ${\cal O}(N)$ iterations to converge!  Similarly, compared to the explicit method our scheme is ncfl ($\sim$ 10-1000) times faster.
Our method should be easily implemented in parallel using standard parallel linear algebra libraries like ScaLAPACK \cite{bla97}.

Although our method is not monotonicity-preserving for arbitrarily large timesteps, the temperature oscillations are of quite small amplitude 
and are damped with time.
Using test problems we show that large speedups (up to $\sim$100-1000 for our test problems) are  achieved compared to the explicit method, 
without seriously violating the monotonicity constraint. A similar directional splitting may also prove effective for isotropic diffusion. Although ADI is
numerically stable, fast, and second order accurate in time for
isotropic diffusion, it does not give strong damping in the large timestep
limit (just like the Crank Nicolson scheme; e.g., \cite{pre92}).

We also tried unsplit methods, both fully implicit using limiters for
the `transverse' terms, and semi-implicit where only `transverse' terms with
limiters are treated 
explicitly.  The limiters lead to nonlinearities that require some care
with an iterative solver, and these unsplit methods result in a large sparse
matrix equation 
which is much more expensive to solve than the tri-diagonal systems of
the split method, even with an iterative solver like conjugate gradients
or Loose GMRES.
Although both of these unsplit methods with limiters appear to be monotonicity-preserving for arbitrary $\Delta t$, it takes many iterations to
obtain a converged solution in both cases, and we generally find that
the split algorithm is quite efficient by comparison.
Our method might be able to serve as an effective preconditioner to
further accelerate unsplit iterative methods, and works quite well as it is for our
present purposes.

Thermal conduction is primarily along the magnetic field direction for
hot plasmas. For astrophysical plasmas large temperature gradients exist,
and it is important for the numerical scheme implementing anisotropic
conduction to yield positive temperatures in regimes of
interest. Another practical requirement is that  
the scheme be fast so that the conduction timestep is not much smaller than the MHD timestep. Here we present a simple, directionally-split 
method that gets close enough
to monotonicity that the temperature remains positive in presence of
relatively large temperature gradients, and results in a large speedup.  
Our method should find applications in modeling of hot astrophysical plasmas with large temperature gradients (e.g., the multiphase 
interstellar/intracluster medium, transition from chromosphere to corona in the Sun).

\section{Acknowledgements}
PS was supported by NASA through Chandra Postdoctoral Fellowship grant number PF8-90054 awarded by the Chandra X-ray Center, which is 
operated by the Smithsonian Astrophysical Observatory for NASA under
contract NAS8-03060, and GWH was supported at the Princeton Plasma
Physics Laboratory by DOE Contract No. DE-AC02-09CH11466.
This research was supported in part by the National Science Foundation through TeraGrid resources provided by NCSA and Purdue University. 
Some of the runs were carried out on {\em Henyey}, the theoretical astrophysics computing cluster at the University of California, Berkeley.
PS thanks Eliot Quataert for encouragement, and Jim Stone and Ben Chandran for useful discussions. We are grateful to Ian Parrish for discussions 
and for his comments on the paper. We thank the anonymous referees for very thorough referee reports that helped improve the quality of the paper substantially.


\begin{thebibliography}{00}
\bibitem{ale96} V. Alexiades, G. Amiez and P. Gremaud, Super-time-stepping acceleration of explicit schemes for parabolic problems, {\em Communications in Numerical Methods in Engineering} {\bf 12} (1996), pp. 31-42.
\bibitem{Baker05} A. H. Baker, E. R. Jessup and T. Manteuffel, A technique for accelerating the convergence of restarted GMRES, {\em SIAM Journal
  on Matrix Analysis and Applications} {\bf 26} (2005), pp. 962-984.
\bibitem{bal08} D. S. Balsara, D. A. Tilley and C. J. Howk, Simulating anisotropic thermal conduction in supernova remnants -- I. Numerical methods, {\em Monthly Notices of Royal Astronomical Society} {\bf 386} (2008), pp. 627-641.
\bibitem{bas02} P. J. Basser and D. K. Jones, Diffusion-tensor MRI: theory, experimental design and data analysis - a technical review, {\em NMR in Biomedicine} {\bf 15} (2002) pp. 456-467.
\bibitem{ber02} Berkowitz, B., Characterizing flow and transport in fractured geological media: A review, {\em Advances in Water Resources} {\bf 25} (2002), pp. 861-884.
\bibitem{bla97} L. S. Blackford, A. Cleary, J. Choi, E. D'Azevedo, J. Demmel, I. Dhillon, J. Dongarra, S. Hammarling, G. Henry, A. Petitet, K. Stanley, D. Walker and R. C. Whaley, ScaLAPACK Users' Guide, SIAM, Philadelphia, PA (1997).
\bibitem{bra65} S. I. Braginskii, Transport Processes in a Plasma, in: M. A. Leontovich (Ed.), Reviews of Plasma Physics, v. 1,  Consultants Bureau, New York, (1965).
\bibitem{dia91} Z. Dian-Lin, C. Shao-Chun, W. Yun-Ping, L. Li, W. Xue-Mei, X. L. Ma and K. H. Kuo, Anisotropic thermal conductivity of the 2D single quasicrystals: Al$_{65}$Ni$_{20}$Co$_{15}$ and Al$_{62}$Si$_3$Cu$_{20}$Co$_{15}$, {\em Physical Review Letters} {\bf 66} (1991), pp.  2778-2781.
\bibitem{don09} R. Dong and J. M. Stone, Buoyant Bubbles in Intracluster Gas: Effects of Magnetic Fields and Anisotropic Viscosity, {\em Astrophysical Journal} {\bf 704} (2009), pp. 1309-1320.
\bibitem{Gustafsson78} I. Gustafsson, A class of first order factorization methods, {\em BIT} {\bf 18} (1978), pp. 142-156.
\bibitem{ker81} D. S. Kershaw, Differencing of the Diffusion Equation in Lagrangian Hydrodynamic Codes, {\em Journal of Computational Physics}, {\bf 39} (1981), pp. 375-395.
\bibitem{lev02} R. J. Leveque, Finite Volume Methods for Hyperbolic Problems, Cambridge University Press (2002).
\bibitem{Lipnikov10} K. Lipnikov, D. Svyatskiy and Y. Vassilevski, Interpolation-free monotone finite volume method for diffusion equations on polygonal meshes, {\em J. Comp. Phys.}, {\bf 228} (2009), pp. 703.
\bibitem{nor07} I. Nordbotten, I. Aavatsmark and G. T. Eigestad, Monotonicity of control volume methods, {\em Numerische Mathematik}, {\bf 106} (2007), pp. 255-288.
\bibitem{par05} I. J. Parrish and J. M. Stone, Nonlinear Evolution of the Magnetothermal Instability in Two Dimensions, {\em Astrophysical Journal}, {\bf 633} (2005), pp. 334-348.
\bibitem{par09} I. J. Parrish, E. Quataert and P. Sharma, Anisotropic Thermal Conduction and the Cooling Flow Problem in Galaxy Clusters, {\em Astrophysical Journal} {\bf 703} (2009), pp. 96-108.
\bibitem{per90} P. Perona and J. Malik, Scale-space and edge detection using anisotropic diffusion, {\em IEEE Transactions on Pattern Analysis and Machine Intelligence} {\bf 12} (1990), pp. 629-639.
\bibitem{pre92} W. H. Press, S. A. Teukolsky, W. T. Vetterling and B. P. Flannery, Numerical Recipes in C: The Art of Scientific Computing, 
Cambridge Univ. Press (1992). 
\bibitem{ras08} Y. Rasera and B. Chandran, Numerical Simulations of Buoyancy Instabilities in Galaxy Cluster Plasmas with Cosmic Rays and Anisotropic Thermal Conduction, {\em Astrophysical Journal}, {\bf 685} (2008), pp. 105-117.
\bibitem{rus10} M. Ruszkowski and S. P. Oh,  Shaken and stirred: conduction and turbulence in clusters of galaxies, {\em Astrophysical Journal} {\bf 713} (2010), pp. 1332-1342.
\bibitem{saa86} Y. Saad and M. H.  Schultz, GMRES: A generalized minimal residual algorithm for solving nonsymmetric linear systems, {\em SIAM Journal on Scientific and Statistical Computing} {\bf 7} (1986), pp.  856-869.
\bibitem{sha07} P. Sharma and G. W. Hammett, Preserving monotonicity in anisotropic diffusion, {\em Journal of Computational Physics} {\bf 227} (2007), pp. 123-142.
\bibitem{sha09} P. Sharma, P. Colella and D. F. Martin, Numerical Implementation of Streaming Down the Gradient: Application to Fluid Modeling of Cosmic Rays, submitted to {\em SIAM Journal on Scientific Computing}, arXiv0909.5426.
\bibitem{sha10} P. Sharma, I. J. Parrish and E. Quataert, Thermal Instability with Anisotropic Thermal Conduction and Adiabatic Cosmic Rays: Implications for Cold Filaments in Galaxy Clusters, {\em Astrophysical Journal}, {\bf 720} (2010), pp. 652-665.
\bibitem{Sheng09} Z. Sheng, J. Yue and G. Yuan, Monotone Finite Volume Schemes of Nonequilibrium Radiation Diffusion Equations on Distorted Meshes  {\em SIAM J. Sci. Comput.}, {\bf 31} (2009), pp. 2915.
\bibitem{sov04} C. R. Sovinec et. al., Nonlinear magnetohydrodynamics simulation using high-order finite elements, {\em Journal of Computational Physics} {\bf 195} (2004) pp. 355-386.
\bibitem{van77} B. J. van Leer, Towards the Ultimate Conservative Difference Scheme. 
IV. A New Approach to Numerical Convection, {\em Journal of Computational  Physics}, {\bf 23} (1977), pp. 276-299.
\end{thebibliography}
\end{document}